\documentclass{revtex4}

\usepackage{amssymb}

\newcommand{\be}{\begin{eqnarray}}
\newcommand{\ee}{\end{eqnarray}}

\begin {document}

\title {Parameter estimation in nonextensive thermostatistics}
\author {Jan Naudts}
\address{
  Departement Natuurkunde, Universiteit Antwerpen,\\
  Universiteitsplein 1, 2610 Antwerpen, Belgium\\
  E-mail {\tt Jan.Naudts@ua.ac.be}
}
\begin {abstract}
Equilibrium statistical physics is considered from the point of view of statistical estimation theory.
This involves the notions of statistical model, of estimators, and of exponential family.
A useful property of the latter is the existence of identities,
obtained by taking derivatives of the logarithm of the partition sum.
It is shown that these identities still exist for models belonging to
generalised exponential families, in which case they involve escort probability distributions.
The percolation model serves as an example.
A previously known identity is derived. It relates the average number of sites belonging to the finite cluster at
the origin, the average number of perimeter sites, and the derivative of the order parameter.
\end {abstract}

\maketitle


\section {Introduction}

Central to this paper is the notion of escort probability distribution.
It was introduced in non-extensive thermostatistics \cite {TMP98} on heuristic
grounds. The notion has been generalised in \cite {NJ04}, where it was shown
that the definition of exponential family can be generalised by considering pairs of
probability distributions, one of which is called the escort of the other.
The probability distributions of nonextensive thermostatistics then
belong to generalised exponential families.

The example used in the present paper is that of the site percolation problem.
Percolation theory is a well-established domain of statistical physics \cite {SD85}.
The number of publications in this domain is vast. It is not the intention of the
present work to make progress in understanding percolation but rather to show that
the notions of escort probabilities and of generalised exponential family
in a natural way fit in and lead to an interesting result.

The next two sections relate equilibrium statistical physics to estimation theory
and recall well-known material such as the definition of the exponential family.
In Section 4, the definition of escort probability distribution is given.
It is used in subsequent sections to recall the generalised lower bound of Cramer and Rao and the definition
of generalised exponential families. Section 7 shows that for any generalised exponential
family one can derive identities, as many as there are model parameters. 
Sections 8 and 9 treat the example of percolation. The final Section 10 contains
a short discussion.


\section {Parameter estimation in statistical physics}

In first instance, statistical physics studies statistical properties of model systems.
Hence, the mathematical theory of parameter estimation can be applied. In this theory, the average value
of one or more quantities, called estimators, is used to estimate parameters of the model.
Here, these quantities are called Hamiltonians because, quite often in statistical physics, one of the parameters
of the model is inverse temperature $\beta$ and because energy, which is the average value of
the Hamiltonian, is used to estimate temperature.

More formally, a model consists of a probability distribution $p_\theta$, which depends on 
some parameters $\theta_1,\theta_2,\cdots\theta_n$, and a set of Hamiltonians $H_1, H_2, \cdots H_n$,
which can be used to estimate the value of the parameters. In statistical physics all these Hamiltonians
are added up to form a single Hamiltonian $H$. This is not done here for simplicity
of notations. Let us illustrate this point.
The Hamiltonian of the $d=1$-Ising model is
\be
H=-J\sum_{m=1}^{n-1}\sigma_m\sigma_{m+1}-h\sum_{m=1}^n\sigma_m.
\ee
The variables $\sigma_m$ can take on the values $\pm 1$. The probability distribution
of the model is
\be
p(\sigma)=\frac 1{Z}\exp(-\beta H)
\quad\hbox{ with }\quad
Z=\sum_{\sigma}\exp(-\beta H).
\ee
The parameters of the model are inverse temperature $\beta>0$ and external field $h$.
However, it is convenient to introduce new parameters
$\theta_1=\beta$ and $\theta_2=\beta h$, and corresponding Hamiltonians
\be
H_1(\sigma)=-J\sum_{m=1}^{n-1}\sigma_m\sigma_{m+1}
\quad\hbox{ and }\quad
H_2(\sigma)=-\sum_{m=1}^n\sigma_m.
\ee
Then, using Einstein's summation convention, the probability distribution can be written as
\be
p(\sigma)=\frac 1{Z}\exp(-\theta^k H_k(\sigma)).
\ee


\section {Exponential family}

A parametrised probability distribution $p_\theta(i)$ belongs to the exponential family
if it can be written into the form
\be
p_\theta(i)=c(i) e^{G(\theta)-\theta^kH_k(i)}.
\label {expfam}
\ee
The function $G(\theta)$ is determined by the normalisation condition
and is given by
\be
G(\theta)=-\ln\sum_ic(i)e^{-\theta^kH_k(i)}.
\ee

In the present paper, the variables $H_k$ are called Hamiltonians.
At first sight one might think that every non-vanishing probability can be written in
exponential form. However, the variables $c$ and $H_k$ in (\ref {expfam})
must not depend on the parameters $\theta^k$. This poses a rather strong condition,
which is not always satisfied. An example of a distribution not belonging to the
exponential family is
\be
p_\alpha(i)=\frac 1{Z(\alpha)}\frac 1{\alpha^2+i^2}
\quad\hbox{ with }\quad
Z(\alpha)=\sum_{i=0}^\infty\frac 1{\alpha^2+i^2}.
\ee
Of course, the Ising model belongs to the exponential family with two parameters.


\section {Escort distributions}
\label {multifractal}

Given a model with probability distribution $p_\theta(i)$,
any other probability distribution $P_\theta(i)$, depending on the same parameters,
is an {\sl escort} distribution for the given $p_\theta(i)$. However, of interest are pairs of
distributions $p_\theta(i),P_\theta(i)$ which satisfy some special relation. In \cite {NJ04} the well-known
inequality of Cramer and Rao was generalised to pairs of probability distributions and a sufficient
condition was given that, when satisfied, makes the inequality optimal.
The usual inequality of Cramer and Rao is optimal in case of a distribution belonging to the
exponential family. Hence it is natural to say that escort probabilities, optimising the
generalised version of the inequality of Cramer and Rao, generalise the notion
of exponential family. This statement is elaborated in the next section.

The concept of escort probability distributions is borrowed from the theory
of fractals, see \cite {BS93}. It goes back to the thermodynamical
analysis \cite {HJKPS86} of multifractals \cite {BPPV84}, now twenty years ago.

Given a probability distribution $p(i)$ which does not depend on any parameters,
one can construct a parameter-dependent family $p_\theta(i)$ by
\be
p_\theta(i)=\frac 1{\sum_jp(j)^\theta}p(i)^\theta.
\ee
Clearly is $p_1=p$. A short calculation gives
\be
\frac {\partial\,}{\partial\theta}p_\theta=p_\theta\left(\frac {\partial G}{\partial \theta}-H(i)\right)
\label {Pfrac}
\ee
with
\be
G(\theta)=-\ln\sum_ip(i)^\theta
\qquad\hbox{ and }\quad
H(i)=-\ln p(i).
\ee
One can indeed write
\be
p_\theta(i)=\exp(G(\theta)-\theta H(i)).
\ee
This shows that the multifractal model belongs to the exponential family.

In the present terminology, $p_\theta$ is an escort of itself, not of $p$,
as said in \cite {BS93}.
But except for this slight change in the meaning of the word {\sl escort},
the present concept generalises that of the multifractal context. Also 
the thermodynamical formalism, developed in the
theory of multifractals, coincides with that found in \cite {NJ04}.


\section {Generalised inequality of Cramer and Rao}

Here we follow \cite {NJ04}, with changes in presentation.
Introduce the notations
\be
\langle A\rangle_\theta=\sum_ip_\theta(i)A(i)
\qquad\hbox{ and }\quad
\langle\langle A\rangle\rangle_\theta=\sum_iP_\theta(i)A(i).
\ee
The notion of score variables, used in statistics, is generalised to
\be
X_k(i)=\frac 1{P_\theta(i)}\frac{\partial\,}{\partial\theta^k}p_\theta(i).
\ee
Let $H_k$ be variables for which a function $F(\theta)$ exists such that
\be
\langle H_k\rangle_\theta=\frac {\partial\,}{\partial\theta^k}F(\theta).
\ee
Then the generalised inequality of Cramer and Rao, valid for arbitrary $u^k$ and $v^l$, reads
\be
& &
u^ku^l\left[\langle\langle H_kH_l\rangle\rangle_\theta
-\langle\langle H_k\rangle\rangle_\theta\langle\langle H_l\rangle\rangle_\theta\right]
v^mv^n\langle\langle X_mX_n\rangle\rangle_\theta
\ge
\left[u^kv^l\frac {\partial^2\,}{\partial\theta^k\partial\theta^l}F(\theta)\right]^2.\cr
& &
\label {gencramerrao}
\ee
The inequality is said to be optimal if equality holds whenever $u=v$.

A sufficient condition for optimality is that a function $Z(\theta)$ exists which is such that
\be
\frac 1{Z(\theta)}\langle\langle X_kX_l\rangle\rangle_\theta
=Z(\theta)\left[\langle\langle H_kH_l\rangle\rangle_\theta
-\langle\langle H_k\rangle\rangle_\theta\langle\langle H_l\rangle\rangle_\theta\right]
=-\frac {\partial\,}{\partial\theta^k}\langle H_l\rangle_\theta.
\label {sc1}
\ee
An slightly stronger condition is that functions $Z(\theta)>0$ and $G(\theta)$ exist for which
\be
\frac {\partial\,}{\partial\theta^k}p_\theta(i)=Z(\theta)P_\theta(i)\left(\frac {\partial G}{\partial\theta^k}-H_k(i)\right).
\label {sc2}
\ee
See Appendix A. It is the latter condition that is used to define generalised exponential families.


\section {Generalised exponential families}
\label {genexpfam}

Obvious solutions of (\ref {sc2}) are of the form
\be
p_\theta(i)&=&c(i)f_i\bigg(G(\theta)-\theta^kH_k(i)\bigg)
\label {pspecform}\\
P_\theta(i)&=&\frac {c(i)}{Z(\theta)}f'_i\bigg(G(\theta)-\theta^kH_k(i)\bigg)\\
Z(\theta)&=&\sum_ic(i)f_i'\bigg(G(\theta)-\theta^kH_k(i)\bigg),
\ee
where $c(i)$ is a positive constant and
where $f_i(x)$ is a positive non-decreasing (stochastic) function. $f_i'(x)$ is the derivative of $f_i(x)$
(For convenience, the dependence of $f_i(x)$ on the stochastic variable $i$ is written as an index;
when possible the dependence on $i$ is omitted).
The normalisation $G(\theta)$ must be such that
\be
\sum_ic(i)f_i\bigg(G(\theta)-\theta^kH_k(i)\bigg)=1.
\ee

A function $\phi_i(y)$ is defined by
\be
\phi_i(y)=f_i'\bigg(f_i^{-1}(y)\bigg),
\ee
with $f_i^{-1}(y)$ the inverse of the function $f_i(x)$. Then one has
\be
\frac 1{c(i)}P_\theta(i)=\frac 1{Z(\theta)}\phi_i\bigg(\frac 1{c(i)}p_\theta(i)\bigg).
\ee
The only function which is its own derivative is the exponential function.
Hence, with $f_i(x)=\exp(x)$ one finds $\phi(x)=x$, $Z(\theta)=1$, and $P_\theta(i)=p_\theta(i)$.
In this case one recovers a distribution belonging to the exponential family.

The function $f(x)$, when not stochastic, has been called a $\phi$-exponential function \cite {NJ04,NJ04bis} because
it satisfies the equation $f'(x)=\phi(f(x))$. The inverse function is then called a $\phi$-logarithm.
Models with a probability distribution satisfying (\ref {pspecform}) are said to belong
to the $\phi$-exponential family.

\section {Identities}

A well-known trick of statistical physics is the calculation of averages by taking derivatives of
$-\ln Z$. E.g., in the Ising model is
\be
-\frac {\partial \,}{\partial\beta}\ln Z=
\frac {\partial G}{\partial\beta}=\langle H\rangle_\theta
\quad\hbox{ and }\quad
-\frac {\partial \,}{\partial h}\ln Z=
\frac {\partial G}{\partial h}=\langle H_2\rangle_\theta.
\ee
As a consequence, the main problem of equilibrium statistical physics is often the evaluation of
the partition sum $Z$ to a closed form expression.

The trick works for all generalised exponential families, and is based on the identity
\be
\frac {\partial G}{\partial\theta^k}=\langle\langle H_k\rangle\rangle_\theta.
\label {identities}
\ee
The latter follows from conservation of probability. Indeed, one calculates
\be
0&=&\frac {\partial\,}{\partial\theta^k}\sum_ip_\theta(i)\cr
&=&\sum_i
Z(\theta)P_\theta(i)\left(\frac {\partial G}{\partial\theta^k}-H_k(i)\right)\cr
&=&Z(\theta)\left(\frac {\partial G}{\partial\theta^k}-\langle\langle H_k\rangle\rangle_\theta\right).
\ee

It may be disappointing that the quantities that one can calculate via (\ref {identities})
are $\langle\langle H_k\rangle\rangle_\theta$, and not the physically relevant $\langle H_k\rangle_\theta$.
However, the derivatives of the latter are static susceptibilities, up to a sign, and are given by
\be
\frac {\partial\,}{\partial \theta^l}\langle H_k\rangle_\theta
=-Z(\theta)\bigg(\langle\langle H_kH_l\rangle\rangle_\theta
-\langle\langle H_k\rangle\rangle_\theta\,\langle\langle H_l\rangle\rangle_\theta\bigg).
\label {fluct}
\ee
Hence, the static susceptibilities are controlled by the pair correlations of the escort distribution,
showing the physical relevance of the latter. Note that the bracket in the r.h.s.~of (\ref {fluct})
coincides with the generalised Fisher information matrix --- see \cite {NJ04}.

The existence of the identities (\ref {identities}) and (\ref {fluct}) makes it attractive to use
generalised exponential families. This is illustrated below in case of the percolation model.

\section {Site percolation}

In the site percolation problem the points of a lattice are occupied with probability $p$,
independent of each other. The origin of the lattice is either unoccupied, with
probability $p_\emptyset$, or it belongs to a cluster of occupied sites.
This cluster is finite with probability one if the density $p$ is less than the
percolation threshold $p_c$. Two clusters of the same size can have a different shape
(these shapes are called lattice animals).
The probability that the origin belongs to a cluster of shape $i$ is denoted $p(i)$
and is given by
\be
p(i)=c(i)p^{s(i)}(1-p)^{t(i)},
\label {percoprob}
\ee
where $c(i)$ is the number of clusters of shape $i$, $s(i)$ is the number
of sites, and $t(i)$ is the number of perimeter sites. The latter are unoccupied
sites that have at least one site of the cluster as a neighbour. Identify shape 0 with the absence of
a cluster at the origin. Then (\ref {percoprob}) holds with
$c(0)=1$, $s(0)=0$ and $t(0)=1$.

The probability that the origin belongs to an infinite cluster is denoted $p(\infty)$.
It vanishes for $p< p_c$ and is strictly positive for $p>p_c$. It satisfies
\be
p(\infty)+\sum_ip(i)=1.
\ee
For simplicity of notation we convene that $s(\infty)=t(\infty)=0$
instead of the more obvious infinite value.

It is possible to write (\ref {percoprob}) into the form
\be
p_\theta(i)=c(i)\exp\bigg([G(\theta)-\theta H(i)](s(i)+t(i))\bigg)
\ee
with parameter $\theta$ defined by
\be
\theta=\ln\frac p{1-p},
\ee
with Hamiltonian
\be
H(i)=\frac {t(i)}{s(i)+t(i)},
\ee
and with normalisation
\be
G(\theta)&=&-\ln(1+e^{-\theta})=\ln p.
\ee
This suggests the introduction of a stochastic function $f$ defined by
\be
f_i(G)&=&\exp((s(i)+t(i))G)\qquad \hbox{ if }i<\infty\cr
f_\infty(G)&=&1-\sum_{i\not=\infty}c(i)f_i\bigg(G+H(i)\ln(e^G-1)\bigg).
\ee
Introduce further conventions $c(\infty)=1$ and $H(\infty)=0$. Then
the probability distribution is of the form (\ref {pspecform}).
Note that the derivative of $f$ is given by
\be
f'_i(G)&=&(s(i)+t(i))f_i(G)\qquad \hbox{ if }i\not=\infty\\
f'_\infty(G)&=&\frac 1{1-e^{G}}\frac {\partial\,}{\partial\theta}p_\theta(\infty)\bigg|_{G=G(\theta)}.
\ee
The optimal escort probability distribution is therefore given by
\be
P_\theta(i)&=&\frac {1}{Z(\theta)}(s(i)+t(i))p_\theta(i)\qquad\hbox{ if }i\not=\infty\\
P_\theta(\infty)&=&\frac {1}{Z(\theta)}\frac 1{1-p}\frac {\partial\,}{\partial\theta}p_\theta(\infty),
\ee
with appropriate normalisation $Z(\theta)$, given by
\be
Z(\theta)&=&\langle s+t\rangle_\theta+\frac 1{1-p}\frac {\partial\,}{\partial\theta}p_\theta(\infty).
\label {percores1}
\ee

\section {Percolation identities}

The identity one can derive from (\ref {identities}) reads
\be
\langle\langle H\rangle\rangle_\theta
=\frac {{\rm d}G}{{\rm d}\theta}
=1-p.
\label {percores2}
\ee
Using the definition of the Hamiltonian and of the escort probability one finds
\be
\langle\langle H\rangle\rangle_\theta
=\frac 1{Z(\theta)}\langle t\rangle_\theta.
\label {percores3}
\ee
Combining (\ref {percores1}, \ref {percores2}, \ref {percores3}) the identity becomes
\be
\frac {\partial\,}{\partial\theta}p_\theta(\infty)=\langle M\rangle_\theta,
\qquad\hbox{ with }\quad M(i)=pt(i)-(1-p)s(i).
\label {mainpercores} 
\ee
This result is known --- see (44a) of \cite {SD85}. The variable $M$ acts as an order parameter.
Its average value vanishes for $p<p_c$. It is non-zero for $p>p_c$ and diverges as $(p-p_c)^{\beta-1}$ when $p$ decreases
towards $p_c$.

In the present approach the relevant susceptibility is
\be
\chi=-\frac {{\rm d}\,}{{\rm d}\theta}\langle H\rangle_\theta=-\frac {{\rm d}\,}{{\rm d}\theta}\langle \frac {t}{s+t}\rangle_\theta.
\ee
Using (\ref {fluct}) this becomes
\be
\chi&=&Z(\theta)\bigg(\langle\langle H^2\rangle\rangle_\theta
-\langle\langle H\rangle\rangle_\theta^2\bigg)\cr
&=&\langle HM\rangle_\theta\cr
&=&\langle \frac 1{s+t}M^2\rangle_\theta+(1-p)\langle M\rangle_\theta.
\ee
It is expected to diverge as $|p-p_c|^{-\gamma}$ when $p$ approaches the percolation threshold.


\section {Discussion}

In non-extensive thermostatistics average energy is often calculated using the escort probability distribution.
Next, entropy is maximised under the constraint that the escort averaged energy $\langle\langle H\rangle\rangle$ has some given value $U$
--- see \cite {TMP98}. The main point of the present paper is that $\langle\langle H\rangle\rangle$
and $\langle\langle H^2\rangle\rangle$
can be used as part of the identities (\ref {identities}, \ref {fluct}). They provide a convenient way
of obtaining relevant information about the model under study. This approach has been illustrated
by taking site percolation as an example.

The presentation has been purely classical, avoiding quantum statistics. A first attempt to
treat the quantum case is found in \cite {NJ05}. The probability distribution and its escort should
be replaced by a pair of density operators $\rho$ and $\sigma$. However, the possibility that $\rho$
and $\sigma$ do not commute prevents a straightforward translation of the classical formalism to the
quantum context.

{
  \appendix
  
  \setcounter{equation}{0}  
  \section*{Appendix A: Optimality of (\ref {sc1}) and implication of (\ref {sc1}) by (\ref {sc2})}

Let us first show that (\ref {sc1}) implies optimality of inequality (\ref {gencramerrao}).
From (\ref {sc1}) follows
\be
\frac {\partial\,}{\partial\theta^k}\langle H_l\rangle_\theta
=\frac {\partial\,}{\partial\theta^l}\langle H_k\rangle_\theta.
\ee
Hence, a function $F(\theta)$ exists such that
\be
\langle H_k\rangle_\theta=\frac {\partial F}{\partial\theta^k}.
\ee
Using (\ref {sc1}) the inequality then becomes
\be
u^ku^l\frac {\partial^2 F}{\partial\theta^k\partial\theta^l}
v^mv^n\frac {\partial^2 F}{\partial\theta^m\partial\theta^n}
\ge
\left[u^kv^l\frac {\partial^2 F}{\partial\theta^k\partial\theta^l}\right]^2.
\ee
Obviously, this is an equality when $u=v$. Hence, the inequality is satisfied optimally.

Finally, assume (\ref {sc2}) holds and prove (\ref {sc1}).
Condition (\ref {sc2}) implies that
\be
0&=&\sum_i\frac {\partial\,}{\partial\theta^k}p_\theta(i)\cr
&=&Z(\theta)\left(\frac {\partial G}{\partial\theta^k}-\langle\langle H_k\rangle\rangle_\theta\right).
\ee
and
\be
X_k(i)=Z(\theta)\left(\frac {\partial G}{\partial\theta^k}-H_k\right)
=-Z(\theta)\left(H_k-\langle\langle H_k\rangle\rangle_\theta\right).
\ee
Hence, the first equality of (\ref {sc1}) follows.
On the other hand is
\be
\frac {\partial\,}{\partial\theta^k}\langle H_l\rangle_\theta
&=&Z(\theta)\sum_iP_\theta(i)\left(\frac {\partial G}{\partial\theta^k}-H_k\right)H_l(i)\cr
&=&-Z(\theta)\left[\langle\langle H_kH_l\rangle\rangle_\theta
-\langle\langle H_k\rangle\rangle_\theta\langle\langle H_l\rangle\rangle_\theta\right].
\ee
This proves the remaining equality.

} 

\section* {}

\end {document}